\documentclass[journal,draftclsnofoot,onecolumn,12pt]{IEEEtran}
\IEEEoverridecommandlockouts

\usepackage{graphicx}
\usepackage{subfig}
\usepackage{cite}
\usepackage{amsmath}
\usepackage{caption}
\usepackage{amsfonts,amssymb}
\usepackage[aboveskip=10pt,belowskip=-15pt,font=footnotesize]{caption}
\usepackage[CJKbookmarks=true,bookmarksnumbered=true,bookmarksopen=true]{hyperref}

\usepackage{color}

\let\oldcite\cite
\renewcommand{\cite}[1]{\mbox{\oldcite{#1}}}

\ifCLASSINFOpdf
\else
\fi

\begin{document}

\title{Design and Implementation of MIMO Transmission Based on Dual-Polarized Reconfigurable Intelligent Surface}

\author{
\IEEEauthorblockN{Xiangyu Chen, Jun Chen Ke, Wankai Tang, Ming Zheng Chen, Jun Yan Dai,\\
 Ertugrul Basar, Shi Jin, Qiang Cheng, and Tie Jun Cui}

\thanks{X. Chen, W. Tang, and S. Jin are with the National Mobile Communications Research Laboratory, Southeast University, Nanjing 210096, China.
}
\thanks{J. C. Ke, M. Z. Chen, Q. Cheng, and T. J. Cui are with State Key Laboratory of Millimeter Waves, Southeast University, Nanjing 210096, China.
}
\thanks{J. Y. Dai is with the State Key Laboratory of Terahertz and Millimeter Waves, City University of Hong Kong, Hong Kong SAR 999077, China.
}
\thanks{E. Basar is with the CoreLab, Department of Electrical and Electronics Engineering, Koc University, Istanbul 34450, Turkey.
}
}

\maketitle

\begin{abstract}
Multiple-input multiple-output (MIMO) signaling is one of the key technologies of current mobile communication systems. However, the complex and expensive radio frequency (RF) chains have always limited the increase of MIMO scale. In this paper, we propose a MIMO transmission architecture based on a dual-polarized reconfigurable intelligent surface (RIS), which can directly achieve modulation and transmission of multichannel signals without the need for conventional RF chains. Compared with previous works, the proposed architecture can improve the integration of RIS-based transmission systems. A prototype of the dual-polarized RIS-based MIMO transmission system is built and the experimental results confirm the feasibility of the proposed architecture. The dual-polarized RIS-based MIMO transmission architecture provides a promising solution for realizing low-cost ultra-massive MIMO towards future networks.
\end{abstract}


\IEEEpeerreviewmaketitle

\section{Introduction}
Multiple-input multiple-output (MIMO) transmission can significantly increase the channel capacity, spectrum efficiency, and reliability of communication systems\cite{goldsmith2005wireless}. Generally, the larger the scale of the antenna array, the greater the performance gain can be brought by MIMO. This basic principle has lead to the research of massive MIMO and ultra-massive MIMO\cite{larsson2014massive},\cite{han2018ultra}. However, as the scale of the MIMO antenna array increases, the size, power consumption, and hardware cost will also increase, especially the complex and expensive radio frequency (RF) chains. These factors limit the further increase of MIMO scale.

In recent years, reconfigurable intelligent surface (RIS) has attracted great attention in the field of wireless communications due to its electromagnetic (EM) manipulation capabilities\cite{cui2014coding}. An RIS can dynamically modulate EM waves under the control of external signals, such as manipulating the amplitude\cite{zhao2013tunable} or phase\cite{zhao2019programmable} of EM waves. The EM manipulation capabilities of RISs can be exploited to improve the channel quality to enhance the performance of wireless communication systems, thus achieving smart radio environments\cite{huang2019reconfigurable, wu2019intelligent, han2019large, di2020smart, tang2020pathloss }. Furthermore, RISs enable direct modulation of EM waves to realize RF chain-free transmitters. An RIS-based frequency-shift keying (FSK) modulation scheme was proposed in\cite{ zhao2019programmable },\cite{tang2019programmable } and\cite{ henthorn2019direct } realized phase-shift keying (PSK) modulation based on RISs, and\cite{basar2020reconfigurable} proposed using an RIS to implement index modulation. \cite{zhang2021symbiotic} proposed using RIS to implement BPSK backscatter communication and to assist primary MIMO transmission at the same time. Moreover, an RIS-based MIMO transmission system, which exploited the array structure of the RIS, was proposed in\cite{tang2020wireless} and\cite{tang2020mimo}. The RIS-based MIMO transmission system can directly use the RIS to perform the modulation and transmission of multiple signals without the need for RF chains, thus significantly reduces the cost and complexity of MIMO systems. However, RISs used in these prior works on RIS-based transmitters are all single-polarized, and the polarization degree of freedom of RISs has not yet been explored. Recently, \cite{ ke2021linear } proposed a dual-polarized RIS to realize polarization manipulation of EM waves. \cite{ desena2021irsassisted } proposed using a dual-polarized RIS to alleviate the imperfect successive interference cancellation of massive MIMO with non-orthogonal multiple access (NOMA). The unit cells of the dual-polarized RIS can independently regulate the phase of EM waves in two orthogonal polarizations. This feature makes it possible to utilize the polarization degree of freedom of dual-polarized RIS to improve the integration of RIS-based MIMO transmission systems.

In this paper, we propose a general architecture of dual-polarized RIS-based MIMO transmission system, which utilizes a dual-polarized RIS to modulate and transmit multiple signals in two polarizations. A scheme to implement quadrature amplitude modulation (QAM) and MIMO transmission based on the dual-polarized RIS is designed. Further, we build a prototype of a dual-polarized RIS-based 2$\times$2 MIMO-QAM transmission system and conduct performance tests.

\section{System Model}
\begin{figure}
	\centering
	\includegraphics[width=0.9\textwidth]{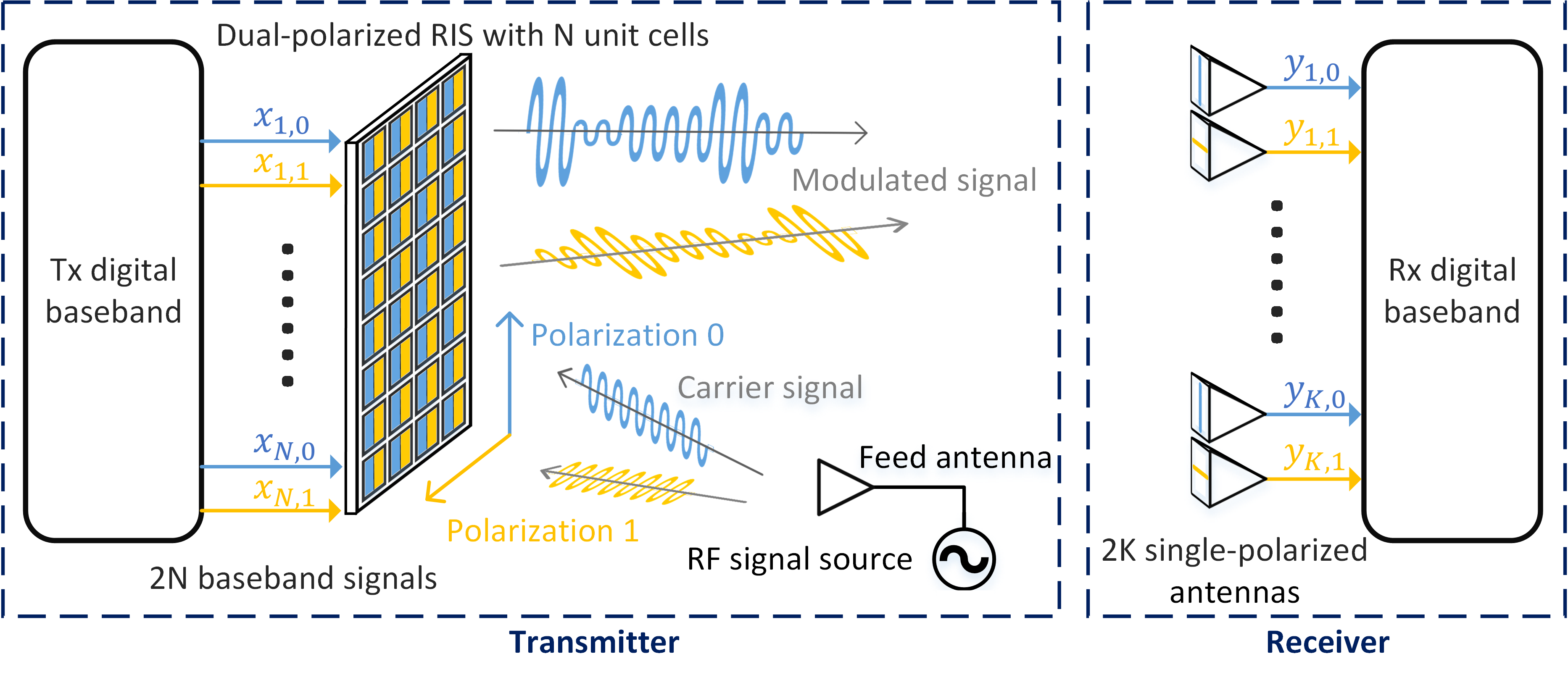}
	\caption{. Dual-polarized RIS-based MIMO transmission system.}
	\label{architecture}
\end{figure}
The general architecture of a dual-polarized RIS-based MIMO transmission system is shown in Fig. \ref{architecture}, in which the dual-polarized RIS is employed to modulate and transmit multiple signals in two polarizations. The transmitter side contains a single-tone carrier signal source, a feed antenna, Tx digital baseband, and a dual-polarized RIS. The dual-polarized RIS consists of $N$ dual-polarized unit cells, each of which can modulate the amplitude and phase of the reflected EM waves in two orthogonal polarizations. The two polarizations are denoted as polarization 0 and polarization 1, respectively. The single-tone carrier signal is transmitted through the feed antenna and radiated on the dual-polarized RIS, and there are carrier components in both polarizations. Each unit cell of the dual-polarized RIS can modulate two baseband signals on the reflected EM waves in polarization 0 and polarization 1, respectively. Thus the dual-polarized RIS can modulate and transmit up to $2N$ signals. There are $K$ single-polarized receiving antennas for each polarization at the receiver side. The received baseband signal vector is denoted as
\begin{equation}\label{equ1}
\mathbf{y}=\left[y_{1,0},\ldots,y_{K,0},y_{1,1},\ldots,y_{K,1}\ \right]^T\in\mathbb{C}^{2K\times1},
\end{equation}
where $y_{k,p}$ is the received signal of the $k^{th}$ receiving antenna in polarization $p$. $\mathbf{y}$ can be expressed as
\begin{equation}\label{equ2}
\mathbf{y}=\sqrt P\mathbf{H}_2\mathbf{\Phi}\mathbf{H}_1\mathbf{c}+\mathbf{w},
\end{equation}
where $P$ is the power of single-tone carrier signal transmitted by the feed antenna, $\mathbf{H}_1\in\mathbb{C}^{2N\times2}$ is the channel from the feed antenna to the dual-polarized RIS, $\mathbf{H}_2\in\mathbb{C}^{2K\times2N}$ is the channel from the dual-polarized RIS to the receiving antennas, the diagonal matrix $\mathbf{\Phi}\in\mathbb{C}^{2N\times2N}$ is the reflection coefficient matrix of the dual-polarized RIS, which also carries the baseband signals, $\mathbf{c}\in\mathbb{C}^{2\times1}$ is the decomposition coefficient of the carrier signal over the two polarizations, and $\mathbf{w}\in\mathbb{C}^{2K\times1}$ is the noise vector of the receiver.

The channel matrix $\mathbf{H}_1$ can be partitioned as
\begin{equation}\label{equ3}
\mathbf{H}_1=\left[\begin{matrix}\mathbf{h}_{1,00}&\mathbf{h}_{1,01}\\\mathbf{h}_{1,10}&\mathbf{h}_{1,11}\\\end{matrix}\right],
\end{equation}
where $\mathbf{h}_{1,pq}\in\mathbb{C}^{N\times1}$ is the channel from the feed antenna in polarization $q$ to the dual-polarized RIS in polarization $p$. Since the feed antenna is usually placed at a fixed location relative to the RIS where it can directly illuminate the RIS to form the transmitter part of the system when it is actually deployed, so $\mathbf{H}_1$ is a deterministic line-of-sight (LoS) channel.

Similarly, $\mathbf{H}_2$ can be partitioned as
\begin{equation}\label{equ4}
\mathbf{H}_2=\left[\begin{matrix}\mathbf{H}_{2,00}&\mathbf{H}_{2,01}\\\mathbf{H}_{2,10}&\mathbf{H}_{2,11}\\\end{matrix}\right],
\end{equation}
where $\mathbf{H}_{2,pq}\!\in\!\mathbb{C}^{K\!\times\!N}$ is the channel from the dual-polarized RIS in polarization\;$q$ to the receiving antenna in polarization\;$p$.

The diagonal matrix $\mathbf{\Phi}$ can be expressed as
\begin{equation}\label{equ5}
\mathbf{\Phi}=\left[\begin{matrix}\mathbf{\Phi}_0&\\&\mathbf{\Phi}_1\\\end{matrix}\right],
\end{equation}
in which
\begin{equation}\label{equ6}
\begin{aligned}
\mathbf{\Phi}_0&=\mathrm{diag}\left\{x_{1,0},...,x_{N,0}\right\}\\
&=\mathrm{diag}\left\{A_{1,0}e^{{j\phi}_{1,0}},\ldots,A_{N,0}e^{{j\phi}_{N,0}}\right\},
\end{aligned}
\end{equation}
\begin{equation}\label{equ7}
\begin{aligned}
\mathbf{\Phi}_1&=\mathrm{diag}\left\{x_{1,1},...,x_{N,1}\right\}\\
&=\mathrm{diag}\left\{A_{1,1}e^{{j\phi}_{1,1}},\ldots,A_{N,1}e^{j\phi_{N,1}}\right\}.
\end{aligned}
\end{equation}

Here, $A_{n,p}$ and $e^{{j\phi}_{n,p}}$ represent the amplitude reflection coefficient and phase shift of the $n^{th}$ unit cell to the incident waves in polarization $p$, respectively, and we have $A_{n,p}\in[0, 1]$, $\phi_{n,p}\in[0,2\pi)$. $x_{n,p}=A_{n,p}e^{{j\phi}_{n,p}}$ is the normalized equivalent baseband signal transmitted by the $n^{th}$ unit cell in polarization $p$, and the baseband signal vector can be written as
\begin{equation}\label{equ8}
\mathbf{x}=\left[x_{1,0},\ldots,x_{N,0},x_{1,1},\ldots,x_{N,1}\right].
\end{equation}

We can rewrite (\ref{equ2}) to show the essence of the dual-polarized RIS-based MIMO transmission system. Since $\mathbf{H}_1$ is a deterministic channel from the feed antenna to the dual-polarized RIS, and the decomposition coefficient $\mathbf{c}$ is a fixed vector that satisfies $||\mathbf{c}||=1$, where $||\cdot||$ denotes the modulus operation, thus
\begin{equation}\label{equ9}
\widetilde{\mathbf{c}}=\mathbf{H}_1\mathbf{c}\in\mathbb{C}^{2N\times1}
\end{equation}
is also a deterministic vector. Denote
\begin{equation}\label{equ10}
\mathbf{E}=\mathrm{diag}\left\{{\widetilde{\mathbf{c}}}_{(1)},{\widetilde{\mathbf{c}}}_{(2)},\ldots,{\widetilde{\mathbf{c}}}_{(2N)}\right\},
\end{equation}
where ${\widetilde{\mathbf{c}}}_{(i)}$ is the $i^{th}$ element of the column vector $\widetilde{\mathbf{c}}$. Then (2) can be rewritten as
\begin{equation}\label{equ11}
\mathbf{y}=\sqrt P\mathbf{H}_2\mathbf{Ex}+\mathbf{w},
\end{equation}
where the normalized baseband signal vector $\mathbf{x}$ is also the reflection coefficient vector of the dual-polarized RIS. The diagonal matrix $\mathbf{E}$ can be regarded as the constant attenuation coefficient matrix of these $2N$ signals caused by $\mathbf{H}_1$ and $\mathbf{c}$.

(\ref{equ11}) reveals the basic principle of our dual-polarized RIS-based MIMO transmission system. The dual-polarized RIS-based MIMO transmitter can directly modulate the information on the carrier by controlling the reflection coefficients of the dual-polarized RIS, so it does not need complex and expensive RF chains as the conventional MIMO transmitter. Moreover, the dual-polarized RIS utilizes the degree of freedom of polarization, where each unit of the dual-polarized RIS can transmit two signals with orthogonal polarizations. Compared with a single-polarized RIS, the dual-polarized RIS with the same number of unit cells can double the number of transmitted signals. Therefore, the dual-polarized RIS-based MIMO transmission scheme can achieve better integrability, thus offering a promising scheme for low-cost and low-complexity MIMO transmission.

\section{Design of MIMO-QAM Transmission Based on Dual-Polarized RIS}
In this section, we first introduce the EM characteristics of the dual-polarized RIS, and then provide a scheme for implementing MIMO-QAM transmission based on the dual-polarized RIS.

\subsection{Dual-Polarized RIS}\label{RIS}
\begin{figure}
\centering
\includegraphics[width=0.8\textwidth]{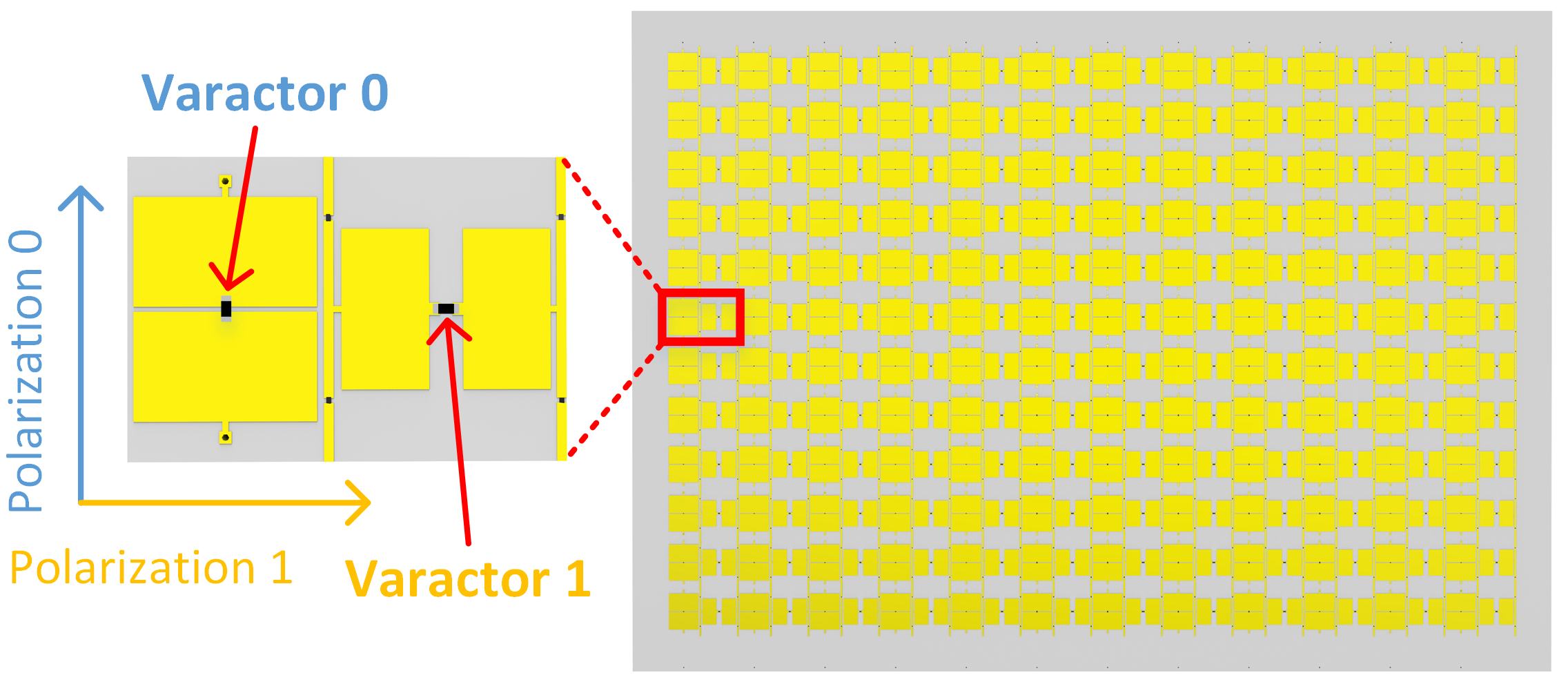}
\caption{. Dual-polarized RIS.}
\label{DualPRIS}
\end{figure}
The dual-polarized RIS is shown in Fig. \ref{DualPRIS}, which can control the phase shift of reflected EM waves in two polarizations\cite{ ke2021linear}. The total length and width of the dual-polarized RIS are 542 mm$\times$410 mm, and the operating frequency is 2.7 GHz. The RIS is composed of 12$\times$12 reduplicated dual-polarized unit cells. Each unit cell has a length and width of 36 mm$\times$25 mm, which is approximately 0.324$\lambda$$\times$0.225$\lambda$ relative to the wavelength $\lambda$. The dual-polarized unit cell consists of the substrate, two pairs of mutually perpendicular rectangular metal patches, two varactor diodes connected across the metal patches, and microstrip lines. For each unit cell, two bias voltages are applied to the two varactor diodes through the two pairs of metal patches. The bias voltages can change the reflection coefficient of the dual-polarized unit for polarization 0 and polarization 1 and thus controls the phase shift of the reflection EM waves. The amplitude reflection efficiency of the unit cell is greater than 70 $\%$, and the amplitude reflection coefficient fluctuates within 1 dB when the bias voltage changes. More details of the dual-polarized RIS can be found in \cite{ke2021linear}. Due to the unique design of the dual-polarized unit cell, the varactor 0 as shown in Fig. \ref{DualPRIS} can control the phase of the reflected polarization 0 EM waves, while the varactor 1 can control the phase of the reflected polarization 1 EM waves.

\begin{figure}
	\centering
	\includegraphics[width=0.65\textwidth]{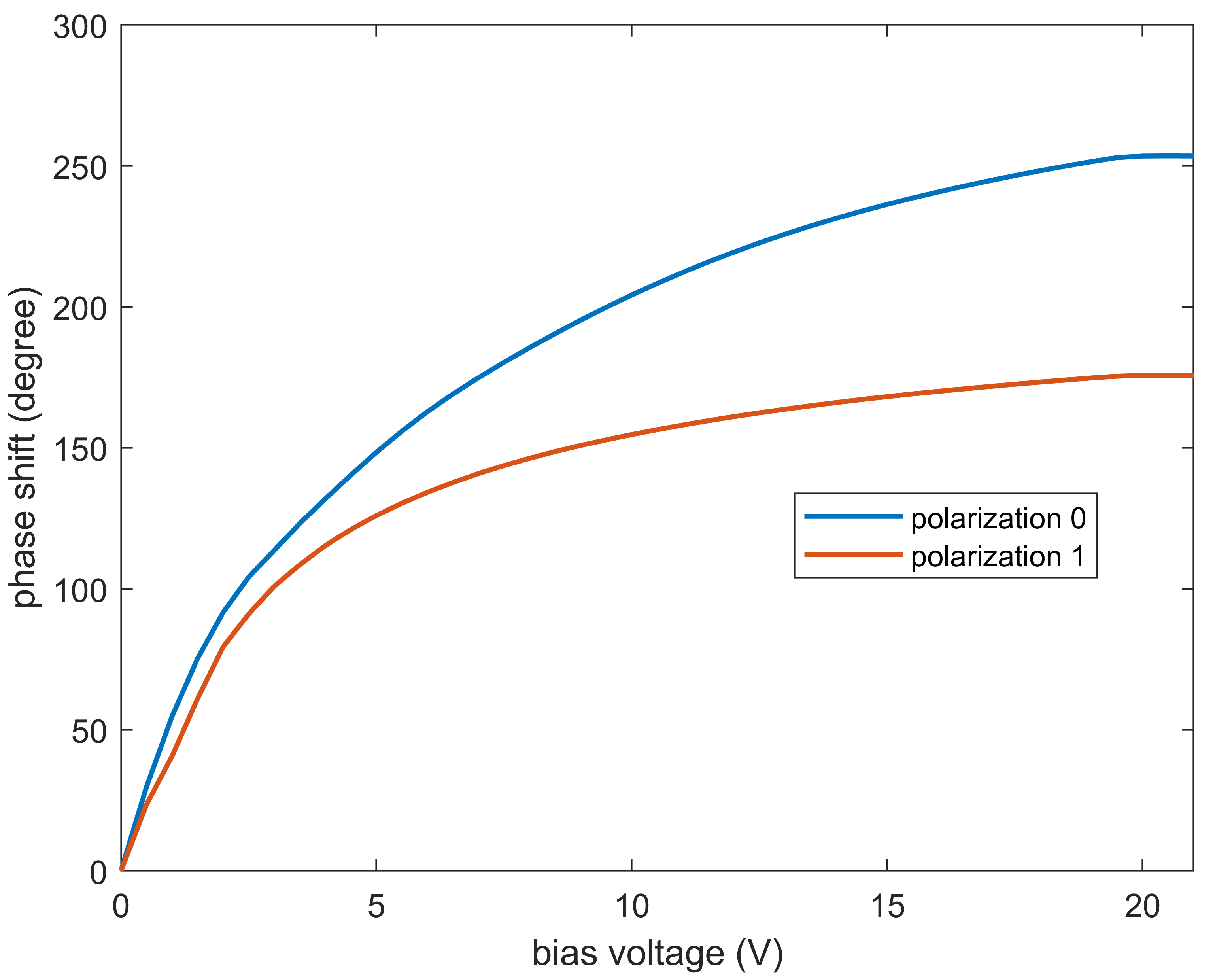}
	\caption{. Measured relationship between the phase shift and the bias voltage of dual-polarized RIS.}
	\label{phasevoltage}
\end{figure}

Fig. \ref{phasevoltage} shows the measured relationship between the phase shift of the reflected EM waves and the bias voltage of the unit cell. It can be seen that the phase shift of the reflected EM waves in both polarizations increases monotonically with the corresponding bias voltage, but the phase shifts of the two polarizations have different relationships with the bias voltage. This is because the dual-polarized unit cell here is not a rotational symmetric structure.

\subsection{Dual-Polarized RIS-based MIMO-QAM Transmission}\label{QAM}
The dual-polarized RIS is phase-controllable and it is difficult to directly realize QAM due to the limitation of constant envelope. Therefore, we use the non-linear modulation technique proposed in\cite{dai2018independent} and\cite{dai2020realization }.

The baseband symbol for non-linear modulation is
\begin{equation}\label{equ12}
{x_{n,p}}\left( t \right){\rm{ = }}\left\{ {\begin{array}{*{20}{c}}
{{e^{j\frac{{\Delta {\varphi _{n,p}}}}{{{T_s}}}({T_s} - {t_{n,p}} - t)}},}&{t \in [0,{T_s} - {t_{n,p}}],}\\
{{e^{j\frac{{\Delta {\varphi _{n,p}}}}{{{T_s}}}(2{T_s} - {t_{n,p}} - t)}},}&{t \in ({T_s} - {t_{n,p}},{T_s}],}
\end{array}} \right.
\end{equation}
where $T_s$ is the symbol period, $t_{n,p}\in[0,T_s]$ is the cyclic time shift, and the phase of $x_{n,p}\left(t\right)$ decreases linearly with the slope of $\frac{\Delta\varphi_{n,p}}{T_s}$.

The baseband symbol in (\ref{equ12}) will generate the $-1^{st}$ order harmonic with frequency $f_c-\frac{1}{T_s}$, and the amplitude $a_{n,p}^{-1}$ and phase $\varphi_{n,p}^{-1}$ of the $-1^{st}$ order harmonic can be written as\cite{dai2020realization }
\begin{equation}\label{equ13}
a_{n,p}^{-1}=\left|{\mathrm{sinc}\left({\frac{{\Delta{\varphi _{n,p}}}}{2}-\pi}\right)}\right|,
\end{equation}
\begin{equation}\label{equ14}
\begin{aligned}
\varphi _{n,p}^{ - 1} =&  - \frac{{2\pi {t_{n,p}}}}{{{T_s}}} + \frac{{\Delta {\varphi _{n,p}}}}{2} +
\varepsilon (2\pi  - \Delta {\varphi _{n,p}}) \cdot \pi\\
&+\mathrm{mod}(\left\lfloor {\frac{{\Delta {\varphi _{n,p}}}}{{2\pi }} - 1} \right\rfloor ,2) \cdot \pi - \pi,
\end{aligned}
\end{equation}
where $|\cdot|$, $\mathrm{sinc}(\cdot)$, $\mathrm{mod}(\cdot)$, $\left\lfloor\cdot\right\rfloor$, $\varepsilon(\cdot)$ denote the absolute value function, sinc function, remainder function, downward rounding function and step function, respectively.

It can be seen from (\ref{equ13}) and (\ref{equ14}) that we can control $\Delta\varphi_{n,p}$ and $t_{n,p}$ to achieve simultaneous manipulation of the amplitude and phase of the $-1^{st}$ order harmonic, and therefore realize QAM on the $-1^{st}$ order harmonic.

The equivalent baseband signal of non-linear modulation can be denoted as
\begin{equation}\label{equ15}
s_{n,p}=a_{n,p}^{-1}e^{j\varphi_{n,p}^{-1}},
\end{equation}
and the center frequency of the equivalent baseband signal is $f_c-\frac{1}{T_s}$. The corresponding baseband signal vector is
\begin{equation}\label{equ16}
\mathbf{s}=\left[s_{1,0},\ldots,s_{N,0},s_{1,1},\ldots,s_{N,1}\right].
\end{equation}

By replacing $\mathbf{x}$ in (\ref{equ11}) with $\mathbf{s}$, the signal model at the $-1^{st}$ order harmonic frequency point can be written as
\begin{equation}\label{equ17}
\mathbf{y}=\sqrt P\mathbf{H}_2\mathbf{Es}+\mathbf{w}.
\end{equation}

Thus the dual-polarized RIS-based MIMO-QAM transmission can be realized according to (\ref{equ17}).

\section{Implementation of MIMO-QAM Transmission Based on Dual-Polarized RIS}\label{Implementation}
This section presents the implementation and test of our 2$\times$2 MIMO system based on the dual-polarized RIS. The architecture and prototype setup of the dual-polarized RIS-based 2$\times$2 MIMO-QAM transmission system are described in Subsection \uppercase\expandafter{\romannumeral4} A. In Subsection \uppercase\expandafter{\romannumeral4} B, tests of signal coupling and bit error rate (BER) of the prototype are discussed.

\subsection{Prototype Setup}\label{Prototype}
\begin{figure}
\centering
\includegraphics[width=0.9\textwidth]{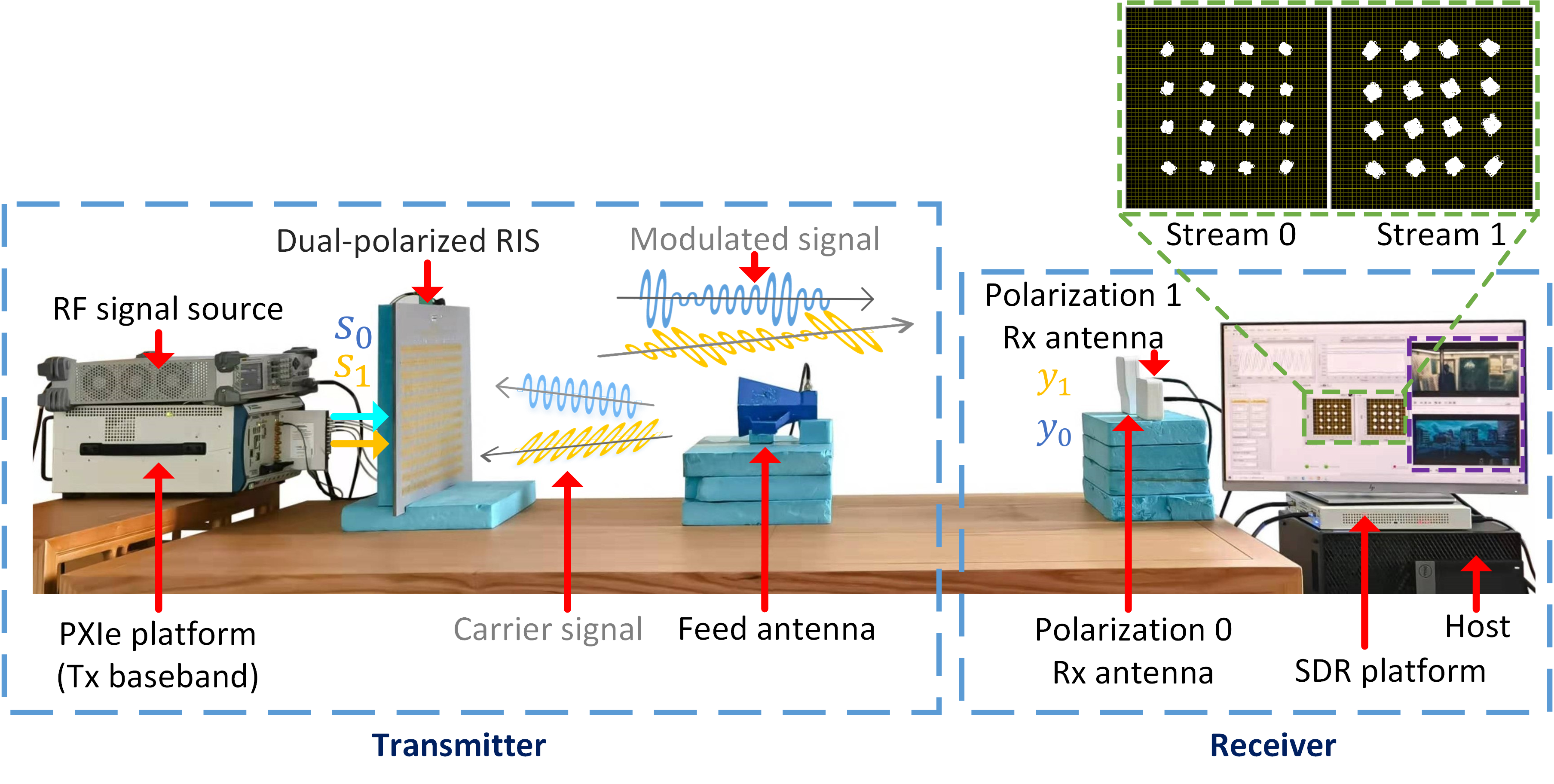}
\caption{. Dual-polarized RIS-based 2$\times$2 MIMO transmission prototype.}
\label{prototype}
\end{figure}
To verify the feasibility of using the dual-polarized RIS to achieve MIMO transmission, we design a 2$\times$2 MIMO communication system based on the dual-polarized RIS, as shown in Fig. \ref{prototype}. The dual-polarized RIS transmits one stream of signal in each of the two polarizations, that is, $s_{1,0},\ldots,s_{N,0}=s_0$ and $s_{1,1},\ldots,s_{N,1}=s_1$. The receiver side uses two single-polarized antennas to receive the polarization 0 signal $y_0$ and the polarization 1 signal $y_1$, respectively.

It should be pointed out that, although due to hardware limitations, the dual-polarized RIS-based MIMO communication system we have implemented can only transmit two data streams, it is enough to verify the feasibility of using the dual-polarization characteristics of RIS to achieve MIMO transmission. When the hardware conditions are met, the system can be expanded according to (\ref{equ11}) and (\ref{equ17}) to realize the transmission of more streams.

Fig. \ref{prototype} presents the prototype of the dual-polarized RIS-based 2$\times$2 MIMO transmission system. The PXIe platform at the transmitter side generates two video bitstreams, namely stream 0 and stream 1, which are mapped into 16-QAM non-linear modulation baseband signals $s_0$ and $s_1$. Then $s_0$ and $s_1$ are converted into analog bias voltage signals by the digital-to-analog converters (DACs) of the PXIe platform and loaded on varactors 0 and varactors 1, respectively. The dual-polarized RIS modulates the two baseband signals on the $-1^{st}$ harmonic of the reflected EM waves in polarization 0 and polarization 1, respectively. The RF signal source generates a 2.7 GHz single-tone carrier signal, which is transmitted by the feed antenna. The feed antenna is a line-polarized horn antenna placed at an angle of 45$^{\circ}$ so that both polarization 0 and polarization 1 have carrier components. The distance between the feed antenna and the RIS is 0.8m, and the distance between the RIS and the Rx antennas is 1.6m. The receiver side utilizes two single-polarized antennas to receive the signals in polarization 0 and polarization 1, then the corresponding digital baseband signals $y_0$ and $y_1$ can be obtained after down-conversion and sampling by software-defined radio (SDR) platform. Subsequent baseband signal processing and demodulation are implemented on the host computer.

\subsection{Experimental Results}\label{Results}
As shown in the top right of Fig. \ref{prototype}, the system can recover the correct constellation and realize the transmission of two video streams. The transmission rate of the system can reach 20 Mbps when the symbol rate is 2.5 MSps. This experimental result verifies the feasibility of MIMO transmission based on a dual-polarized RIS. In particular, it can be seen in Fig. \ref{prototype} that the constellation points are distorted, which we found to be caused by the coupling of two voltage signals on the dual-polarized RIS.

\begin{figure}
\centering
\includegraphics[width=0.85\textwidth]{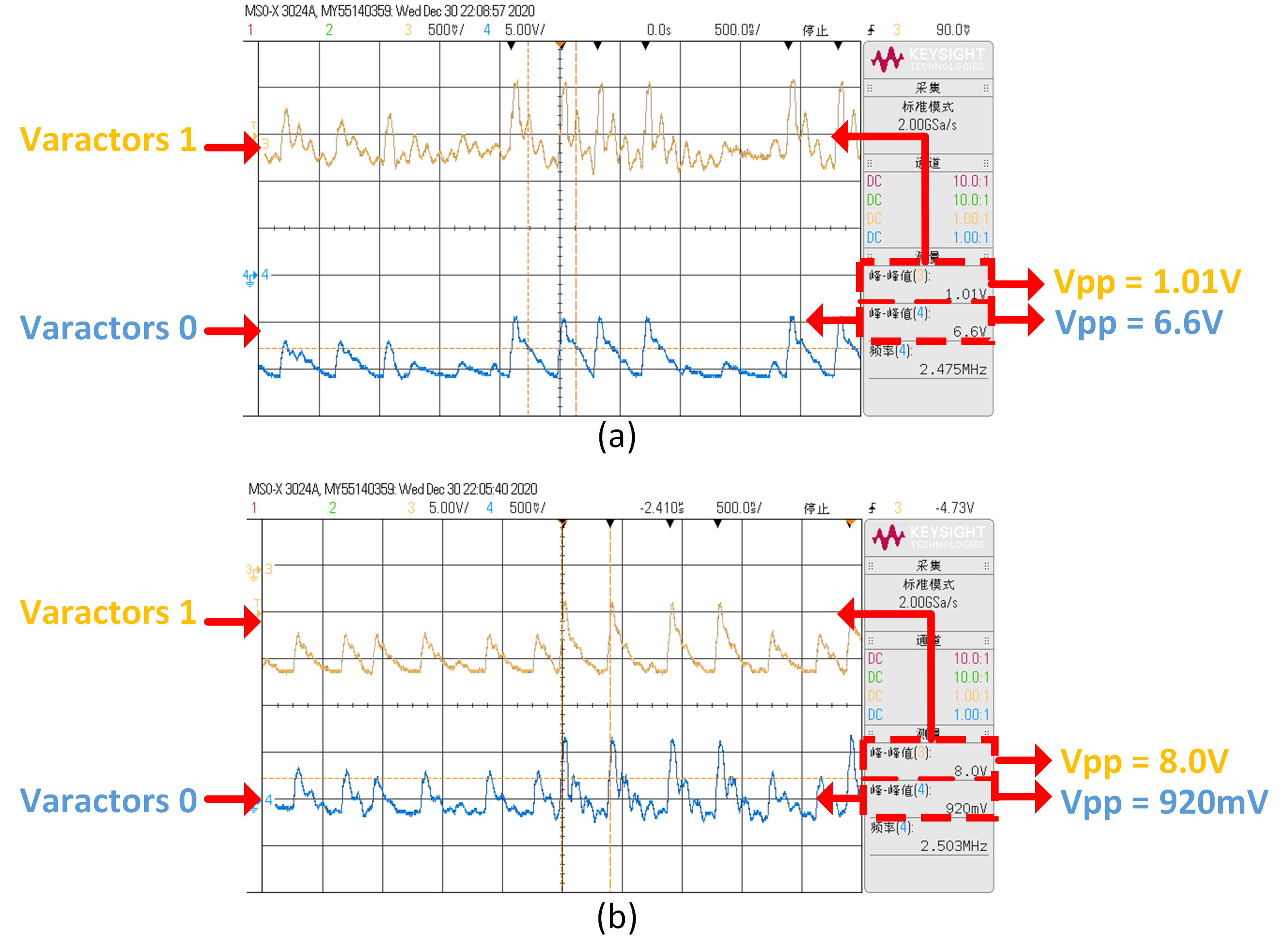}
\caption{. Measurement of voltage signals on the dual-polarized RIS. (a)$s_0$ is turned on and $s_1$ is grounded. (b)$s_0$ is grounded and $s_1$ is turned on.\newline}
\label{coupling}
\end{figure}

\begin{figure}
\centering
\includegraphics[width=0.85\textwidth]{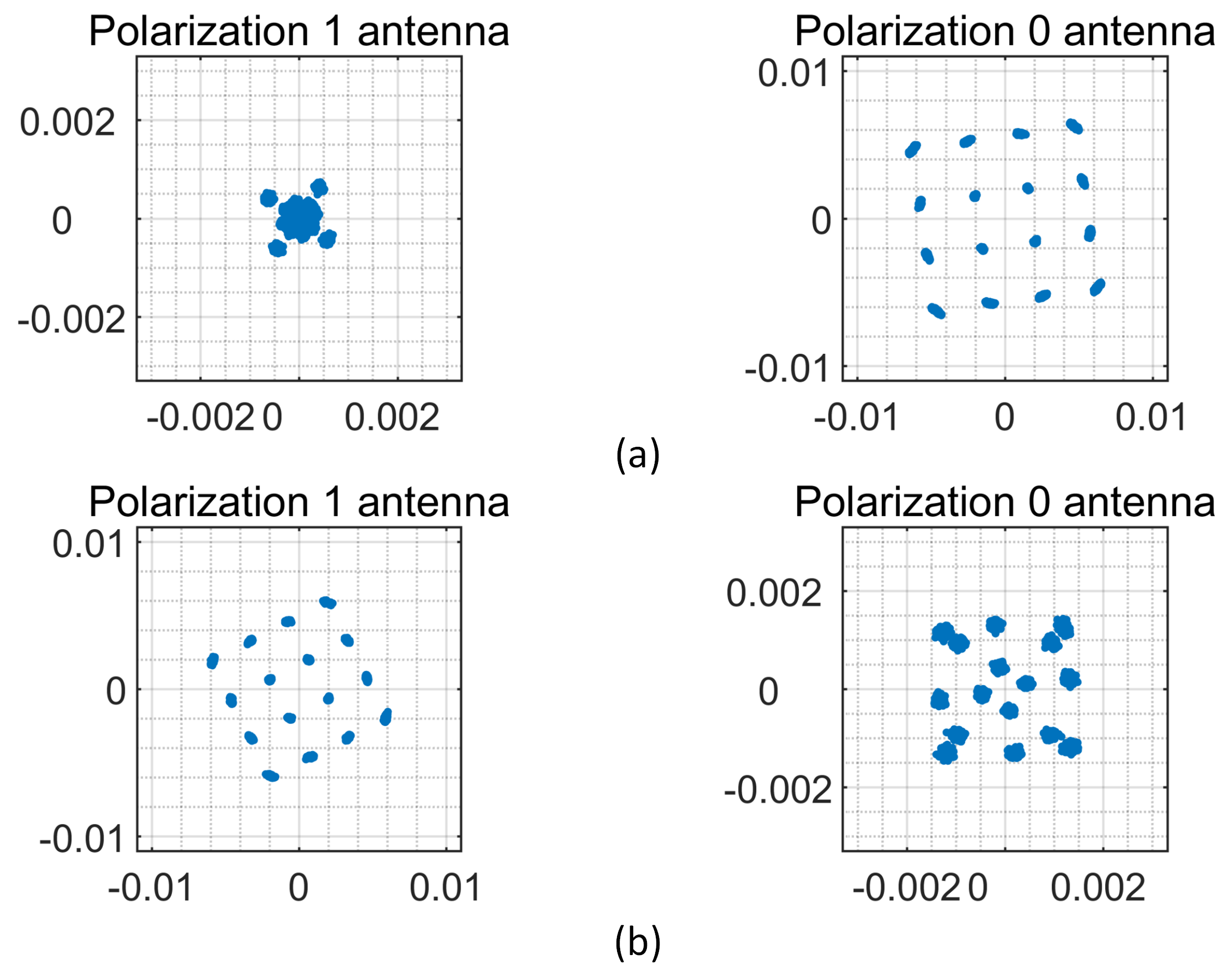}
\caption{. Unequalized signals received by antennas in two polarizations. (a)$s_0$ is turned on and $s_1$ is grounded. (b)$s_0$ is grounded and $s_1$ is turned on.}
\label{unequal}
\end{figure}

We conduct a preliminary measurement of the coupled signals on the dual-polarized RIS. We measure the voltage signals on the varactors directly through an oscilloscope with only one of $s_0$ and $s_1$ is turned on. The symbol rate is 2.5 MSps during the measurement, and the results are shown in Fig. \ref{coupling}. The blue line of the oscilloscope is the common voltage signal on varactors 0, and the yellow line is the common voltage signal on varactors 1. $s_0$ is the blue line in Fig. \ref{coupling}(a), $s_1$ is the yellow line in Fig. \ref{coupling}(b). As can be seen from Fig. \ref{coupling}(a), when $s_0$ is turned on and $s_1$ is grounded, there is still a voltage signal on varactors 1, which is coupled from $s_0$ on varactors 0. We place the receiving antennas close to the dual-polarized RIS to directly observe the unequalized signals in both polarizations, as shown in Fig. \ref{unequal}(a). The QAM constellation points of polarization 0 maintain the correct relative position. But the QAM constellation of polarization 1 is distorted due to the voltage signal on varactors 1 is coupled from the voltage signal on varactors 0. The phenomenon in Fig. \ref{coupling}(b) and Fig. \ref{unequal}(b) is the same. The signal coupling phenomenon is caused by the fact that the dual-polarized RIS is not specifically designed for polarization isolation during the layout process. It can be measured that the isolation degree between varactors 0 and varactors 1 of the dual-polarized RIS is approximately 16 dB (20$\log_{10}$(6.6V/1.01V)$\approx$16dB). Therefore, when both $s_0$ and $s_1$ are turned on, the two voltage signals will be coupled to each other on the dual-polarized RIS causing distortions in the constellation points.

\begin{figure}
\centering
\includegraphics[width=0.75\textwidth]{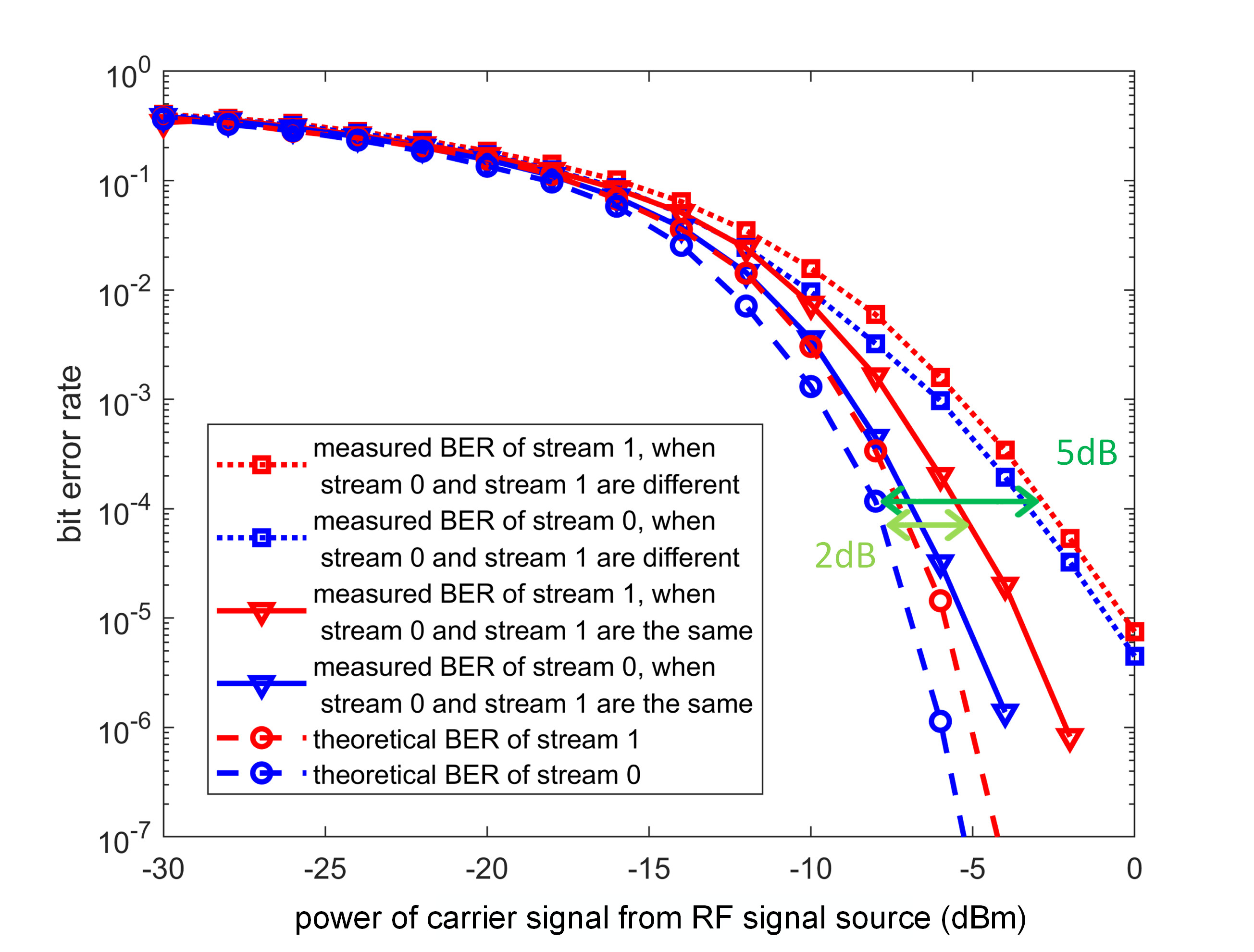}
\caption{. Measured and theoretical BER performances of the prototype.}
\label{BER}
\end{figure}

In order to analyze the impact of the coupling of the two voltage signals, we measured the BER performance of the prototype, as shown in Fig. \ref{BER}. The measurements were performed at a transmission rate of 20 Mbps. The curves of theoretical BER are obtained by substituting the signal-to-noise ratio (SNR) of the received signal into the BER calculation formula of 16-QAM in an additive white Gaussian noise (AWGN) channel\cite{tang2020realization}. It can be seen that the measured BER performance is worse than the theoretical BER. The reason is that the SNR of the received signal we used to calculate the theoretical BER did not take into account the interference of the coupled signal, and the actual SNR of the received signal is lower. When the data streams are the same, that is $s_0=s_1$, the effect of interference is reduced and the measured BER is closer to the theoretical BER without considering coupling. At the BER level of $10^{-4}$, the prototype has a 5 dB loss in BER performance compared to the theoretical value when transmitting two different data streams, and the loss in BER performance is reduced to 2 dB as the interference decreases when transmitting the same data streams. Obviously, the signal coupling on the dual-polarized RIS will cause the degradation of communication performance.

In brief, our prototype verifies the feasibility of MIMO transmission based on a dual-polarized RIS, and we also find that the polarization isolation of a dual-polarized RIS is an important factor. We believe that the polarization isolation of dual-polarized RISs can be improved in the future, like the current well-established dual-polarized antenna arrays.

\section{Conclusion}

In this paper, we have proposed a general architecture of a dual-polarized RIS-based MIMO transmission system and built a 2$\times$2 MIMO-QAM prototype based on the dual-polarized RIS. The prototype achieved real-time transmission of two video streams, which verifies the feasibility of the proposed approach. We also found that signal coupling of the two polarizations on the dual-polarized RIS will cause system performance degradation, implying that polarization isolation should be a key consideration for dual-polarized RIS design. Overall, the proposed dual-polarized RIS-based MIMO transmission architecture is a promising scheme to achieve low-cost ultra-massive MIMO transmission towards future networks.


\bibliographystyle{IEEEtran}
\bibliography{IEEEabrv,mylib}

\begin{thebibliography}{10}
\providecommand{\url}[1]{#1}
\csname url@samestyle\endcsname
\providecommand{\newblock}{\relax}
\providecommand{\bibinfo}[2]{#2}
\providecommand{\BIBentrySTDinterwordspacing}{\spaceskip=0pt\relax}
\providecommand{\BIBentryALTinterwordstretchfactor}{4}
\providecommand{\BIBentryALTinterwordspacing}{\spaceskip=\fontdimen2\font plus
\BIBentryALTinterwordstretchfactor\fontdimen3\font minus
  \fontdimen4\font\relax}
\providecommand{\BIBforeignlanguage}[2]{{%
\expandafter\ifx\csname l@#1\endcsname\relax
\typeout{** WARNING: IEEEtran.bst: No hyphenation pattern has been}%
\typeout{** loaded for the language `#1'. Using the pattern for}%
\typeout{** the default language instead.}%
\else
\language=\csname l@#1\endcsname
\fi
#2}}
\providecommand{\BIBdecl}{\relax}
\BIBdecl

\bibitem{goldsmith2005wireless}
A.~Goldsmith, \emph{Wireless Communications}.\hskip 1em plus 0.5em minus
  0.4em\relax {Cambridge, U.K.: Cambridge University Press}, 2005.

\bibitem{larsson2014massive}
E.~Larsson \emph{et~al.}, ``Massive {MIMO} for next generation wireless
  systems,'' \emph{IEEE Commun. Mag.}, vol.~52, no.~2, pp. 186--195, Feb. 2014.

\bibitem{han2018ultra}
C.~Han, J.~M. Jornet, and I.~Akyildiz, ``Ultra-massive {MIMO} channel modeling
  for graphene-enabled terahertz-band communications,'' in \emph{Proc. IEEE VTC
  Spring}, Jun. 2018, pp. 1--5.

\bibitem{cui2014coding}
T.~J. Cui, M.~Q. Qi, X.~Wan, J.~Zhao, and Q.~Cheng, ``Coding metamaterials,
  digital metamaterials and programmable metamaterials,'' \emph{Light-Sci.
  Appl.}, vol.~3, no.~10, p. e218, Oct. 2014.

\bibitem{zhao2013tunable}
J.~Zhao \emph{et~al.}, ``A tunable metamaterial absorber using varactor
  diodes,'' \emph{New J. Phys.}, vol.~15, no.~4, p. 043049, Apr. 2013.

\bibitem{zhao2019programmable}
J.~Zhao \emph{et~al.}, ``Programmable time-domain digital-coding metasurface
  for non-linear harmonic manipulation and new wireless communication
  systems,'' \emph{Natl. Sci. Rev.}, vol.~6, no.~2, pp. 231--238, Mar. 2019.

\bibitem{huang2019reconfigurable}
C.~Huang \emph{et~al.}, ``Reconfigurable intelligent surfaces for energy
  efficiency in wireless communication,'' \emph{IEEE Trans. Wireless Commun.},
  vol.~18, no.~8, pp. 4157--4170, Aug. 2019.

\bibitem{wu2019intelligent}
Q.~Wu and R.~Zhang, ``Intelligent reflecting surface enhanced wireless network
  via joint active and passive beamforming,'' \emph{IEEE Trans. Wireless
  Commun.}, vol.~18, no.~11, pp. 5394--5409, Nov. 2019.

\bibitem{han2019large}
Y.~Han, W.~Tang, S.~Jin, C.-K. Wen, and X.~Ma, ``Large intelligent
  surface-assisted wireless communication exploiting statistical {CSI},''
  \emph{IEEE Trans. Veh. Technol.}, vol.~68, no.~8, pp. 8238--8242, Aug. 2019.

\bibitem{di2020smart}
D.~Renzo \emph{et~al.}, ``Smart radio environments empowered by reconfigurable
  intelligent surfaces: How it works, state of research, and the road ahead,''
  \emph{IEEE J. Sel. Areas Commun.}, vol.~38, no.~11, pp. 2450--2525, Nov.
  2020.

\bibitem{tang2020pathloss}
W.~Tang \emph{et~al.}, ``Wireless communications with reconfigurable
  intelligent surface: Path loss modeling and experimental measurement,''
  \emph{IEEE Trans. Wireless Commun.}, vol.~20, no.~1, pp. 421--439, Jan. 2021.

\bibitem{tang2019programmable}
W.~Tang \emph{et~al.}, ``Wireless communications with programmable metasurface:
  Transceiver design and experimental results,'' \emph{China Commun.}, vol.~16,
  no.~5, pp. 46--61, May. 2019.

\bibitem{henthorn2019direct}
S.~Henthorn, K.~L. Ford, and T.~. OFarrell, ``Direct antenna modulation for
  high-order phase shift keying,'' \emph{IEEE Trans. Antennas Propag.},
  vol.~68, no.~1, pp. 111--120, Jan. 2020.

\bibitem{basar2020reconfigurable}
E.~Basar, ``Reconfigurable intelligent surface-based index modulation: A new
  beyond {MIMO} paradigm for {6G},'' \emph{IEEE Trans. on Commun.}, vol.~68,
  no.~5, pp. 3187--3196, May. 2020.

\bibitem{zhang2021symbiotic}
Q.~Zhang, Y.~C. Liang, and H.~Vincent~Poor, ``Reconfigurable intelligent
  surface assisted {MIMO} symbiotic radio networks,'' \emph{IEEE Trans. on
  Commun.}, (early access), Mar. 2021.

\bibitem{tang2020wireless}
W.~Tang \emph{et~al.}, ``Wireless communications with programmable metasurface:
  New paradigms, opportunities, and challenges on transceiver design,''
  \emph{IEEE Wireless Commun.}, vol.~27, no.~2, pp. 180--187, Apr. 2020.

\bibitem{tang2020mimo}
W.~Tang \emph{et~al.}, ``{MIMO} transmission through reconfigurable intelligent
  surface: System design, analysis, and implementation,'' \emph{IEEE J. Sel.
  Areas Commun.}, vol.~38, no.~11, pp. 2683--2699, Nov. 2020.

\bibitem{ke2021linear}
J.~C. Ke \emph{et~al.}, ``Linear and nonlinear polarization syntheses and their
  programmable controls based on anisotropic time-domain digital coding
  metasurface,'' \emph{Small Structures}, p. 2000060, Jan. 2021.

\bibitem{desena2021irsassisted}
A.~S. de~Sena \emph{et~al.}, ``{IRS}-assisted massive {MIMO-NOMA} networks:
  Exploiting wave polarization,'' \emph{arXiv:2012.03639}, May. 2021, [online]
  Available: https://arxiv.org/abs/2012.03639.

\bibitem{dai2018independent}
J.~Y. Dai, J.~Zhao, Q.~Cheng, and T.~J. Cui, ``Independent control of harmonic
  amplitudes and phases via a time-domain digital coding metasurface,''
  \emph{Light-Sci. Appl.}, vol.~7, no.~1, pp. 1--10, Nov. 2018.

\bibitem{dai2020realization}
J.~Y. {Dai} \emph{et~al.}, ``Realization of multi-modulation schemes for
  wireless communication by time-domain digital coding metasurface,''
  \emph{IEEE Trans. Antennas Propag.}, vol.~68, no.~3, pp. 1618--1627, Mar.
  2020.

\bibitem{tang2020realization}
W.~Tang \emph{et~al.}, ``Realization of reconfigurable intelligent
  surface-based {Alamouti} space-time transmission,'' in \emph{Proc. IEEE Int.
  Conf. Wireless Commun. Signal Process}, Oct. 2020, pp. 904--909.

\end{thebibliography}

\end{document}